\documentstyle[prl,aps,preprint]{revtex}

\begin{document}
\draft
\title{Geometric phase effects for wavepacket revivals}
\author{C. Jarzynski}
\address{Institute for Nuclear Theory, University of Washington\\
Seattle, WA~~98195}
\date{\today}
\maketitle

\begin{abstract}

The study of wavepacket revivals is extended
to the case of Hamiltonians which are
made time-dependent through the adiabatic
cycling of some parameters.
It is shown that the quantal geometric phase
({\it Berry's phase}) causes the revived packet
to be displaced along the classical trajectory, by
an amount equal to the classical geometric phase
({\it Hannay's angle}), in one degree of freedom.
A physical example illustrating this effect
in three degrees of freedom is mentioned.

\end{abstract}


Ordinarily, a quantal wavepacket following a classical
trajectory spreads irreversibly, until it completely
loses its integrity as a localized packet.
However, when the underlying classical trajectory
is stable and periodic, the dispersed
wavepacket eventually ``puts itself together again'' and
for a time continues, localized, along the classical
trajectory.
The argument for this comes from an examination of the
dynamical phases $\exp -iE_nt/\hbar$ acquired by the
energy eigenstates whose superposition makes up
the wavepacket.\cite{parker86,avper,nau}
Such {\it wavepacket revivals} --- which have been
observed experimentally in Rydberg atoms\cite{sciam}
as well as one-atom masers\cite{rempe} --- are intriguing because
they represent the seemingly spontaneous resurrection of
classical behaviour in a quantal system.
The study of wavepacket revivals has so far been
restricted to systems where the Hamiltonian is
time-independent.
The present paper extends the theory
to the case of slowly time-dependent Hamiltonians,
where, as will be shown, Berry's phase
adds a new and interesting twist to the picture.

Specifically, in this paper I consider wavepacket evolution
under an {\it adiabatically cycled} Hamiltonian, that is,
a parameter-dependent Hamiltonian $\hat H({\bf R})$,
where the parameter $\bf R$ is made to slowly trace out a
closed loop in parameter space.
I will restrict myself to systems of one degree of
freedom, although parameter space is multi-dimensional.
Thus the underlying classical motion (assumed bounded)
is periodic for fixed $\bf R$.
When $\bf R$ is made to change slowly with time,
the classical motion is {\it nearly} periodic:
a trajectory goes round and round a
slowly changing closed curve
in the two-dimensional phase space.
At any instant, this curve is an {\it energy shell} ---
a level surface --- of the instantaneous Hamiltonian,
and is determined by the requirement
that the action $\oint p\,dq$
over one ``period'' is an invariant (see e.g.
Ref.\cite{hannay} for details).
The analysis that follows shows that, as
in the time-independent case, there
will be a revival of the wavepacket:
the standard argument for revivals
is easily extended to the case of a slowly
time-dependent Hamiltonian, only the dynamical phases
now have the form $\exp -(i/\hbar)\int E_n(t)\, dt$.
However, this is only part of the story.
As demonstrated by Berry\cite{berry},
an energy eigenstate evolving under an adiabatically
cycled Hamiltonian acquires not only a
dynamical phase, but also a {\it geometric}
phase, determined by the loop traced out
in ${\bf R}$-space.
A proper analysis of wavepacket evolution in this
situation must take this phase into account.
I show that, if the time at which the parameter
returns to its initial value is chosen to coincide
with the revival time, then the net effect of Berry's
phase is to cause a {\it displacement} of the location
at which the revived wavepacket appears.
Moreover, the amount of this displacement is
given by Hannay's angle\cite{hannay},
the classical analogue of Berry's phase.
As the analysis will show, this effect is generic in
one degree of freedom; at the end of the paper
I briefly mention
a physical example in three degrees of freedom
which illustrates the same effect.

To demonstrate what has been stated above,
it is useful to first review a result
concerning wavepacket evolution under a
time-{\it independent} Hamiltonian $\hat H$;
this result is summarized
by Eq.\ref{eq:prelimresult} below.
Imagine a quantal wavepacket $\psi_0$
localized around some point $(q_0,p_0)$ in
phase space.
(That is, the configuration-space
representation is localized around $q_0$,
the momentum-space representation around $p_0$.
More formally, one could
represent the quantal state directly in
phase space, using a Wigner transform.)
Expressing $\psi_0$ as a superposition
of energy eigenstates $\vert u_n\rangle$, the
non-vanishing coefficients $a_n$
are confined to some finite range of values of $n$,
centered, say, around $\bar n$.
Assume the semiclassical
regime: $\bar n\gg 1$.
Under time evolution, the
coefficients in this superposition acquire
phases:
$\psi(t)=\sum_n \exp(i\alpha_n)
a_n \vert u_n\rangle$,
where $\alpha_n=-E_nt/\hbar$.
It is the process of ``de-phasing'' ---
the fact that the $\alpha_n$'s grow
at different rates --- which determines
how the wavepacket moves and spreads.
The Correspondence Principle, however,
dictates that in the semiclassical limit
($\bar n\rightarrow\infty$),
the packet is propelled along a classical trajectory.
Let us consider this more closely.

A Taylor expansion of $E_n$ around $\bar n$ yields:
\begin{equation}
\label{eq:expansion}
\alpha_n\,=\,-{1\over\hbar} E_{\bar n} t\,-\,
\Delta n{1\over\hbar}E^\prime_{\bar n} t\,-\,
{1\over 2\hbar}(\Delta n)^2
E^{\prime\prime}_{\bar n} t
\,-\,\cdots,
\end{equation}
where $\Delta n\equiv n-\bar n$, and
$E^\prime_{\bar n}\equiv dE_{\bar n}/d{\bar n}$,
(and similarly for the second derivative
$E^{\prime\prime}_{\bar n}$, and higher
derivatives $E^{(s)}_{\bar n}$\cite{btu}.)
Using $1/\bar n$ as an ordering parameter,
$E^{(s)}_{\bar n}$ is
${\it O}(1/\bar n^s)$,
and $\hbar$ is ${\it O}(1/\bar n)$
(in comparison with the relevant ``macroscopic''
quantities $E_{\bar n}$ and
$I_{\bar n}\equiv (1/2\pi)\oint p\,dq$, with the latter
calculated for a classical orbit of energy
$E_{\bar n}$).
The first term on the right side
of Eq.\ref{eq:expansion}, $-E_{\bar n}t/\hbar$
(which scales as $\bar n$), contributes
an external phase to the wavepacket.
The second term ($\sim\bar n^0$) describes the {\it linear}
(in $\Delta n$) de-phasing between the components
$\vert u_n\rangle$ of $\psi_0$,
the third term ($\sim 1/\bar n$) the {\it quadratic} de-phasing,
and so forth.

In the semiclassical limit,
in which the wavepacket follows a classical
trajectory, the first two terms dominate:
\begin{eqnarray}
\psi(t)\,&\cong&\,
\exp(-iE_{\bar n}t/\hbar)\,
\sum_n\exp(-i\Delta n E^\prime_{\bar n}t/\hbar)
a_n\vert u_n\rangle\nonumber\\
\label{eq:alphabeta}
&\equiv&\,{\rm e}^{i\alpha_{\bar n}}
\sum_n{\rm e}^{i\beta_n} a_n\vert u_n\rangle.
\end{eqnarray}
Thus, disregarding the external phase $\alpha_{\bar n}$,
the phases
\begin{equation}
\beta_n\,=\,-\Delta n E^\prime_{\bar n}t/\hbar
\end{equation}
specify a wavepacket that evolves
along a classical trajectory.
(Note in particular that this evolution is periodic.)
This result is valid for times of order unity:
$t\sim\bar n^0$
(by which is meant that the ratio of $t$ to the
relevant macroscopic time scale --- the period
of an orbit of energy $E_{\bar n}$ --- is held
fixed as $\bar n\rightarrow\infty$).
At longer times,
the quadratic and higher-order de-phasing terms
become appreciable, and are
responsible for the spreading of the packet.
Now, WKB theory gives us
\begin{equation}
\label {eq:wkb}
{1\over\hbar}E^\prime_{\bar n}\,\approx\,
{\partial H\over\partial I},
\end{equation}
where $\partial H/\partial I$ is the derivative
of the classical Hamiltonian with respect to the
classical action, evaluated
at energy $E_{\bar n}$.
The correction to Eq.\ref{eq:wkb} is
${\it O}(1/{\bar n}^2)$ (the
${\it O}(1/{\bar n})$ correction is identically zero).
In terms of {\it action-angle variables}
$(\theta,I)$ \cite{goldstein},
$\partial H/\partial I$ is simply the rate of change
of the angle variable $\theta$ along a classical
trajectory.
Thus, neglecting the ${\it O}(1/{\bar n}^2)$ correction
to Eq.\ref{eq:wkb}, we get
\begin{equation}
\beta_n\,=\,-\Delta n\Delta\theta,
\end{equation}
where $\Delta\theta=(\partial H/\partial I)t$
is the change in $\theta$ associated
with classical evolution for time $t$.
We therefore conclude that, if we modify a
semiclassical wavepacket $\psi_0$
by tacking on linear de-phasing factors
$\exp(-i\Delta n\Delta\theta)$ to the terms in
the superposition,
then the effect is to shift the location of
the wavepacket along a classical trajectory,
by a change in the angle variable
equal to the amount of (linear) de-phasing, $\Delta\theta$.
Symbolically,
\begin{equation}
\label{eq:prelimresult}
\Bigl\{
a_n\rightarrow{\rm e}^{-i\Delta n\Delta\theta}a_n
\Bigr\}
\qquad\Longrightarrow\qquad
(\theta,I)\rightarrow(\theta+\Delta\theta,I).
\end{equation}

Let us now move on to adiabatically driven systems.
Consider a parameter-dependent Hamiltonian
$\hat H({\bf R})$, and imagine again
an initial wavepacket $\psi_0$
localized at $(q_0,p_0)$.
Now, however, let the packet evolve under
the time-{\it dependent} Hamiltonian
obtained by making
${\bf R}$ slowly trace out a closed loop
$\Gamma$ in parameter space.
Denote the initial (and final) point on this
loop by ${\bf R}_0$, and the initial and
final times by $t=0$ and $t=T_\Gamma$.

Assuming $d{\bf R}/dt$ slow enough that the quantal
adiabatic theorem holds, the wavefunction at time $t$ is
\begin{equation}
\psi(t)\,=\,
\sum_n{\rm e}^{i\phi_n(t)}a_n
\vert u_n({\bf R}(t))\rangle,
\end{equation}
where the $\phi_n$'s are real, the coefficients
$a_n$ are time-independent, and the
$\vert u_n({\bf R})\rangle$'s
are the eigenstates of $\hat H({\bf R})$.
Thus, when ${\bf R}(t)$ returns to ${\bf R}_0$ at
time $T_\Gamma$, we have a state identical
to $\psi_0$, except that the
expansion coefficients have acquired
phases:
$a_n\rightarrow{\rm e}^{i\phi_n(T_\Gamma)}a_n$.

Before the discovery of Berry's
phase\cite{berry}, one might have guessed that
the phases $\phi_n(T_\Gamma)$ are given by
\begin{equation}
\label{eq:dynamical}
\phi_n(T_\Gamma)\,=\,-{1\over\hbar}
\int_0^{T_\Gamma} dt\,
E_n(t),
\end{equation}
where $E_n(t)$ is the $n$th eigenstate
of $\hat H({\bf R}(t))$.
For the time being, assume Eq.\ref{eq:dynamical}
is correct.
Expanding as in Eq.\ref{eq:expansion}
above, we get
\begin{equation}
\label{eq:tdepexpansion}
\phi_n(T_\Gamma)\,=\,
-{b_0(T_\Gamma)\over\hbar}\,-\,
\Delta n
{b_1(T_\Gamma)\over\hbar}\,-\,
(\Delta n)^2
{b_2(T_\Gamma)\over 2\hbar}\,-\,\cdots,
\end{equation}
where
\begin{equation}
b_s(T_\Gamma)\,=\,\int_0^{T_\Gamma} dt\,
E^{(s)}_{\bar n}(t).
\end{equation}

The sizes of the
terms in Eq.\ref{eq:tdepexpansion} are determined
by a competition between the largeness
of $T_\Gamma$ and the smallness of $1/{\bar n}$.
For instance, suppose we take ${\bar n}\rightarrow\infty$,
while letting $T_\Gamma$ scale as unity:
$T_\Gamma\sim\bar n^0$ (in the sense defined
earlier).
This is the limit in which classical behavior
persists over the entire time of observation.
On the other hand, in this limit
$b_s(T_\Gamma)\sim 1/{\bar n}^s$, so
the quadratic and higher-order de-phasing terms
in Eq.\ref{eq:tdepexpansion} are negligible.
Thus, as in the time-independent case,
we associate linear de-phasing with classical
propagation, and the higher-order terms with spreading.

The regime of interest in this paper
is that for which $T_\Gamma$ is large enough that
the first {\it three}
terms of Eq.\ref{eq:tdepexpansion} are significant,
while terms of order $(\Delta n)^3$ and higher
are not.
(Formally: $\bar n\rightarrow\infty$,
$T_\Gamma\sim\bar n$.)
In this regime the spreading of the
wavepacket becomes important,
and wavepacket {\it revivals} appear.

As mentioned, a revival occurs
when a packet --- after having lost its localized
structure --- reassembles itself,
continues for a while along a classical
trajectory, then again spreads and dissolves.
(Such revivals repeat themselves regularly,
in a pattern of ``quantum beats''\cite{parker86}.
Additionally, there are {\it fractional} revivals\cite{sciam},
but these will not be discussed here.)
To show that revivals appear in
adiabatically cycled systems,
suppose we choose $T_\Gamma$ so that
$b_2(T_\Gamma)/2\hbar=2\pi$.
Then at time $T_\Gamma$ the effect of the quadratic
de-phasing term in Eq.\ref{eq:tdepexpansion} will be
null, and the wavefunction will
reflect only the external phase
$\exp(i\phi_{\bar n})$
and the linear de-phasing term
\begin{equation}
\label{eq:linph}
-\Delta n{b_1(T_\Gamma)\over\hbar}\,
=\,
-\Delta n\int_0^{T_\Gamma}dt\,
{\partial H\over\partial I}(t).
\end{equation}
(We have discarded a correction that scales
like $1/{\bar n}$.)
Here $\partial H/\partial I$ at time $t$ is evaluated
at the energy shell of $H({\bf R}(t))$ which
corresponds to the eigenstate $\bar n$,
in other words at constant $I$, even as $H$ slowly changes.
According to  Eq.\ref{eq:prelimresult}, Eq.\ref{eq:linph}
describes
a wavepacket which has shifted from its original
position in phase space, $(q_0,p_0)$, along a classical
trajectory of $\hat H({\bf R}_0)$,
by a change in the angle variable given by
\begin{equation}
\label{eq:classdynshift}
\Delta\theta\,=\,\int_0^{T_\Gamma}dt\,
{\partial H\over\partial I}(t).
\end{equation}

Eq.\ref{eq:classdynshift} seems to place the revived
wave packet at the point in phase space which the
classical trajectory would have reached after time
$T_\Gamma$.
That is, since $\dot\theta=\partial H/\partial I$
in the time-independent case,
and since in the adiabatic limit the action
$I$ is an invariant\cite{arnold},
it appears at first glance evident that the classical change in
$\theta$ after time $T_\Gamma$ should be given by
Eq.\ref{eq:classdynshift}, with $\partial H/\partial I$
evaluated at constant $I$.
This picture, however, is erroneous:
as demonstrated by Hannay\cite{hannay},
the classical trajectory
experiences an additional shift in $\theta$:
\begin{equation}
\label{eq:hannaysangle}
\Delta\theta\,=\,
\int_0^{T_\Gamma} dt\,
{\partial H\over\partial I}(t)\,+\,
\Delta\theta_H.
\end{equation}
This ``extra'' shift $\Delta\theta_H$,
{\it Hannay's angle}, is {\it geometric}:
it is determined by the
loop $\Gamma$ in parameter space.
I will now show that the source of the discrepancy
--- the reason Hannay's angle did not appear
in Eq.\ref{eq:classdynshift} --- is the neglect
of Berry's phase
in Eq.\ref{eq:dynamical}.

In Ref.\cite{berry}, Berry showed that the correct
phase acquired by an adiabatically cycled
eigenstate consists of both the dynamical phase
of Eq.\ref{eq:dynamical}, and a geometric phase
$\gamma_n$ (determined by the loop $\Gamma$):
\begin{equation}
\phi_n(T_\Gamma)\,=\,-{1\over\hbar}
\int_0^{T_\Gamma}
dt\,E_n(t)\quad +\quad
\gamma_n.
\end{equation}
Calculating the effect of Berry's phase
on the evolution of a wavepacket launched from
$(q_0,p_0)$ is simple.
First, Eq.\ref{eq:tdepexpansion}
gathers the extra terms
\begin{equation}
\label{eq:gamexp}
\gamma_{\bar n}\,+\,
\Delta n{d\gamma_{\bar n}\over d{\bar n}}\,+\,
{1\over 2}(\Delta n)^2
{d^2\gamma_{\bar n}\over d{\bar n}^2}\,+\,\cdots.
\end{equation}
Now, $\gamma_{\bar n}\sim\bar n$
(see e.g. Eq.[30] of Ref.\cite{berry85}),
so $d^s\gamma_{\bar n}/d{\bar n}^s\sim{\bar n}^{1-s}$.
Thus, in the limit $\bar n\rightarrow\infty$
only the first two terms
in Eq.\ref{eq:gamexp} survive.
Let us again choose $T_\Gamma$ so that
$b_2(T_\Gamma)/2\hbar=2\pi$.
In the notation of Eq.\ref{eq:alphabeta},
the wavepacket at time $T_\Gamma$ is then given by:
\begin{equation}
\label{eq:revival}
\psi(T_\Gamma)\,=\,
{\rm e}^{i\alpha_{\bar n}}
\sum_n{\rm e}^{i\beta_n} a_n\vert u_n\rangle,
\end{equation}
where now
\begin{eqnarray}
\alpha_{\bar n}\,&=&\,
-{1\over\hbar}\int_0^{T_\Gamma}
dt\,E_{\bar n}(t)\quad +\quad
\gamma_{\bar n}\\
\label{eq:betar}
\beta_n\,&=&\,
-\Delta n\Bigl[
\int_0^{T_\Gamma} dt\,
{\partial H\over\partial I}(t)
\quad -\quad{d\gamma_{\bar n}\over d{\bar n}}\Bigr].
\end{eqnarray}
This describes a wavepacket whose
position in phase space is specified by both
the ``dynamical'' shift of Eq.\ref{eq:classdynshift},
and an extra shift $-d\gamma_{\bar n}/d{\bar n}$ in the
angle variable:
\begin{equation}
\label{eq:final}
\Delta\theta\,=\,\int_0^{T_\Gamma}
dt\,{\partial H\over\partial I}(t)
\quad -\quad{d\gamma_{\bar n}\over d{\bar n}}.
\end{equation}
To complete the connection with the classical
result, Eq.\ref{eq:hannaysangle},
we invoke the central result of
Ref.\cite{berry85}, which states that
Hannay's angle $\Delta\theta_H$ and Berry's
phase $\gamma_n$ are related semiclassically by:
\begin{equation}
\label{eq:berrysresult}
\Delta\theta_H=-{d\gamma_n\over dn}.
\end{equation}
Thus, the revival
described by Eqs.\ref{eq:revival} - \ref{eq:betar} does
indeed appear where a proper classical analysis suggests
it ought to.
Specifically, the dynamical quantal phases
$\int E_n dt$
are responsible for the ``dynamical'' shift in
the angle variable (Eq.\ref{eq:classdynshift}), while
the geometric quantal phases $\gamma_n$ further
boost the packet by an amount equal to the classical
geometric shift $\Delta\theta_H$.

(A word about the sizes of terms in
Eq.\ref{eq:final}.
Although the first term on the
right side scales like $\bar n$, whereas the
second term is order unity,
the two nevertheless have a comparable effect
on the location of the wavepacket, since
for this purpose their values are only
relevant {\it modulo} $2\pi$.
On the other hand, the leading-order correction
to the first term scales like $1/\bar n$
--- since corrections to Eq.\ref{eq:wkb}
are ${\it O}(1/\bar n^2)$ ---
thus its effect is genuinely small.)

Eq.\ref{eq:final} embodies the central result of
this paper; combined with Eq.\ref{eq:berrysresult},
it reveals the effect of Berry's phase on
wavepacket revivals in adiabatically cycled systems.
In addition to this result,
the formalism used in this paper provides a
simple interpretation of the semiclassical
relationship between Berry's phase and
Hannay's angle (Eq.\ref{eq:berrysresult}).
Namely, if we modify a wavepacket
$\psi_0=\sum_n a_n\,\vert u_n({\bf R}_0)\rangle$
by tacking on Berry's phases
$\exp\,i\gamma_n$
to the terms in the superposition,
then the effect is to shift the packet along a
classical trajectory of $H({\bf R}_0)$, by a change
in the angle variable equal to Hannay's angle,
$\Delta\theta_H$.
This follows directly from Eqs.\ref{eq:prelimresult}
and \ref{eq:berrysresult}, without any need for a
discussion of revivals.
(Alternatively, one could stand the argument on its
head and {\it derive} Eq.\ref{eq:berrysresult} by
considering the semiclassical evolution of a
wavepacket under $\hat H({\bf R}(t))$ and
invoking Eq.\ref{eq:prelimresult}.)
This ``wavepacket interpretation''
of Eq.\ref{eq:berrysresult} is similar in
spirit to arguments presented by Hannay\cite{hannay};
its novelty resides in that it offers an
easy visualization of the relationship between the
two geometric quantities, Berry's phase (quantal) and
Hannay's angle (classical).

It has been assumed in the preceding analysis
that the value of $T_\Gamma$ is chosen
so that a revival occurs just as ${\bf R}(t)$
concludes its circuit in parameter space.
Suppose now that we are less restrictive with
the value of $T_\Gamma$, but still within the
regime of wavepacket revivals (i.e.
$b_2(T_\Gamma)/2\hbar$ is order unity,
but not necessarily $2\pi$).
A straightforward calculation reveals that a
revival then occurs (along the classical trajectory)
at time $T_R$ satisfying
$b_2(T_R)/2\hbar=2\pi$, even though the Hamiltonian
at $t=T_R$ is different from that at $t=0$.
Thus, revivals are as generic in
adiabatically driven systems as in time-independent
ones.
However, only when the revival time $T_R$
coincides with the cycling time $T_\Gamma$ is it
meaningful to discuss the effect of Berry's
phase on the revived wavepacket, since Berry's
phase (as well as Hannay's angle) is well-defined
only for {\it closed} circuits in parameter
space.

The effect of Berry's phase on the revival
location of wavepackets has been illustrated here
in one degree of freedom.
It would of course be very desirable to find a
physical (three-dimensional) system which exhibits
this effect.
One candidate, suggested in conversation by
M.\ Nauenberg,
is a Rydberg atom in a weak magnetic field {\bf B};
here the classical motion is characterized by the
precession of the Kepler orbit at the Larmor
frequency, and with a proper choice of
$\vert{\bf B}\vert$ a wavepacket launched upon
the orbit will experience a revival\cite{nau}.
Now suppose that, upon launching the packet,
we let the {\it direction} of {\bf B}
adiabatically trace out a loop which encloses
a solid angle $\Omega$ in {\bf B}-space, so that
at the revival time the field is back to its initial
orientation.
Then as a result of Berry's phase, the
revived wavepacket appears at a location which
differs from where it would have appeared
had {\bf B} remained constant.
The difference is simply a rotation
by $\Omega$\cite{note} around the initial direction of
{\bf B}.\cite{formalism}
This is essentially an atomic version of the
Foucault pendulum, and if experimentally
observable\cite{new}, would constitute a vivid
demonstration of the geometric phase effects
illustrated in the present paper.

\section*{ACKNOWLEDGEMENTS}

I would like to thank Mark Mallalieu, Jim Morehead,
Michael Nauenberg, and Steve Tomsovic for extended
discussions which clarified crucial issues during
the preparation of this manuscript.
This work was supported by the Department of
Energy under Grant No. DE-FG06-90ER40561.

\end{document}